# Polarization-dependent resonant inelastic X-ray scattering study at the Cu *L* and O *K*-edge of YBa$_2$Cu$_3$O$_{7-x}$

Martin Magnuson[1*], Thorsten Schmitt[2] and Laurent Duda[3]

[1]*Department of Physics, Chemistry and Biology, IFM, Thin Film Physics Division,*
*Linköping University, SE-58183 Linköping, Sweden*
[2]*Research Department Synchrotron Radiation and Nanotechnology, Paul Scherrer Institut,*
*Swiss Light Source (SLS), CH-5232 Villigen PSI, Switzerland*
[3]*Department of Physics and Astronomy, Division of Molecular and Condensed Matter Physics,*
*Uppsala University, Box 516, S-751 20 Uppsala, Sweden*

*Corresponding author: Martin.Magnuson@ifm.liu.se

**Abstract**
We present a study on the high-$T_c$ superconductor (HTSC) YBa$_2$Cu$_3$O$_{7-x}$ (YBCO) using polarization-dependent X-ray absorption and resonant inelastic X-ray scattering. High-resolution measurements using synchrotron-radiation are compared with calculations using a quasi-atomic multiplet approach performed at the Cu $2p_{3/2}$-edge of YBCO. We use a multiplet approach within the single impurity Anderson model to reproduce and understand the character of the localized low-energy excitations in YBCO. We observe a distinct peak at about 0.5 eV in O *K* RIXS. This peak shows dependence on doping, incident energy, and momentum transfer that suggests that it has a different origin than the previously discussed cuprate bi-magnons. Therefore, we assign it to multi-magnon excitations between the Zhang Rice bands and/or the Upper Hubbard bands, respectively.

**Introduction**
YBa$_2$Cu$_3$O$_{7-x}$ (YBCO) is one of the archetypical high $T_c$-superconductors (HTSC) with a transition temperature ($T_c$ = 90.5 K) above the nitrogen boiling point. This is an immense advantage since it makes it easier and more cost effective to cool below its $T_c$, which is a requirement for being able to reap the benefits of using HTSC in technological applications [1]. Naturally, it would be even better if room temperature HTSC could be found. The path to this goal is, however, hampered by the lack of fundamental understanding what drives the metal-superconductor transition (MST) in the first place. Conventional superconductors are driven by the phonon coupling between the charge carriers in the material and the strength of this coupling sets an upper limit to the attainable transition temperature. Intriguingly, HTSCs possess transition temperatures far above the conventional limit within the Bardeen-Cooper-Schrieffer (BCS) theory [2]. Thus the mechanism for high $T_c$-superconductivity remains one of the most nagging problems to be solved in correlated electron physics.

YBCO consists of planes of edge-sharing CuO$_2$ square plaquettes separated by layers containing rare-earth metals and oxygen. YBCO is a unique case for a superconducting cuprate because the *planes* of corner-sharing CuO$_2$ plaquettes are separated by apical oxygens from a different type of layer that consists of one dimensional $CuO_3$ *chains*. The advantage of cuprates is the non-stoichiometry, i.e., continuous tuning of carrier concentration from underdoped to overdoped regions via optimally doped, by adding excess oxygen atoms to reduce oxygen or substitution of metallic atoms. The orthorhombic structure of doped YBCO consists of $CuO_5$ pyramids that are elongated along the height (*c*-axis) as a result of Jahn-Teller lattice distortion and electron correlation effects. A *d*-hole in each $Cu^{2+}$ ion occupies a $dx^2-y^2$ orbital, where the *z*-axis is along the *c*-axis of the crystal. The strong electron correlation effects create localized spins around the Cu sites while the doping and shortening of the *c*-axis leads to the coexistence of the Zhang-Rice spin-singlet multiplet. The Cu atoms are bonded with O atoms in-plane and with apical O. The Cu $dx^2-y^2$ is the highest occupied level of the $CuO_2$ layer.

When the hole-concentration *x* increases through excess of oxygen in the chains, an opposite lattice distortion occurs where the Jahn-Teller elongated $CuO_5$ pyramids contracts along the *c*-axis so that the Cu-apical O distance becomes smaller. The effect of doping thus counteracts the effect of lengthening the *c*-axis by the Jahn-Teller distortion. The balanced equilibrium *c*-axis length caused by the combination of electron correlations and doping, play an important role for the orbitals and multiplets. When a dopant hole is introduced in the valence band of cuprates, it also gives rise to a superposition of singlet hole doped O *2p-σ* orbitals surrounding a Cu $dx^2-y^2$ singlet state forming a localized Zhang-Rice singlet (ZRS) [3] state in the upper Hubbard band of the $CuO_2$ planes. The ZRS is thus a superposition of local singlet states.

Recently, we have presented results on YBCO [4] revealing temperature dependent self-doping effects, which are pivotal for its metal-superconducting transition. Since one-electron band theory does not sufficiently take into account the strong electron correlations in cuprates, we calculated the many-electron states called multiplets in the $CuO_5$ pyramids. We found that the $2p_z$-orbitals of apical oxygen, which hybridize with Cu(1) $3d_{y^2-z^2}$-orbitals in the chains and with Cu(2,3) $3d_{z^2-r^2}$-orbitals in the planes, *reconstruct* upon cooling and an *interlayer charge transfer* from the chains to the planes is induced. This show directly that superconductivity in YBCO is closely linked with and promoted by a self-doping mechanism for providing the planes with the necessary charge density in order to assume an extraordinary high critical temperature $T_c$. However, the cause for this reconstruction still remains an open question. **I**n the present paper**,** we focus on the question whether and how this process is accompanied and possibly mediated by any collective modes, such as phonons and magnons.

## 1. Experimental

### 2.1 Sample preparation

Thin films of $YBa_2Cu_3O_{6+x}$ (YBCO) were grown by pulsed laser deposition on (100) $SrTiO_3$ 5x5x0.5 $mm^3$ single crystal substrates using 25 nm thick $CeO_2$ buffer layers deposited using RF sputtering at a temperature of 780 °C as described elsewhere [4]. The layers were deposited with an RF source power of 100 W in a mixed argon-oxygen atmosphere (60% $O_2$ + 40% Ar) at a partial pressure of 0.1 mbar. The distance between the target and the sample was 30 mm. After $CeO_2$ deposition, the sample was slowly (10°C/min) cooled down to room temperature under 500 mbar oxygen pressure and was transferred into the PLD chamber without breaking vacuum. During the deposition of YBCO, a substrate was placed directly into the sapphire sample holder and heated by a SiC radiation heater from the back side. The YBCO films were deposited at the temperature of 820 °C. The oxygen pressure during deposition was set to 0.6 mbar, the repetition rate to 10 Hz and the laser energy density at the target 1.25 J/cm$^2$. The substrate-to-target distance was set to 60 mm. 7500 laser pulses approximately correspond to YBCO thickness of about 350 nm. To obtain optimal doping, the films were post-annealed in-situ at 550 °C at an oxygen pressure of 750 mbar. The optimally doped sample (x ≈ 0.9) has critical temperature of $T_c$ = 90.5 K and very sharp transition with $\Delta T_c$ ≈ 1 K signifying high quality of the sample.

### 2.1 X-ray emission and absorption measurements

The X-ray absorption (XAS) and resonant inelastic X-ray scattering (RIXS) measurements were performed at the Advanced Resonant Spectroscopies (ADRESS) beamline [5] at the Swiss Light Source (SLS), Paul Scherrer Institut, Switzerland, using the Super-Advanced X-ray Emission Spectrometer (SAXES) [6] [7]. A photon flux of about $10^{13}$ photons/sec/0.01% bandwidth was focused to a spot size below 5x55 μm$^2$ vertically and horizontally (VxH). The Cu $L_3$ edge XAS measurements were made in the bulk-sensitive total fluorescence yield (TFY) mode with an energy resolution better than 100 meV. For the RIXS measurements at the Cu $L_3$ edge (930 eV), the combined energy resolution was 120 meV. To investigate the anisotropy of the electronic structure, linearly polarized X-rays were used at different grazing incidence angles with the electric field vector (**E**) oriented either (nearly) parallel to the *c*-axis (out-of-plane) or parallel to the CuO$_2$ planes (in-plane) defined by the *a*-axis and the *b*-axis. More details of the measurements and sample preparation is described elsewhere [4].

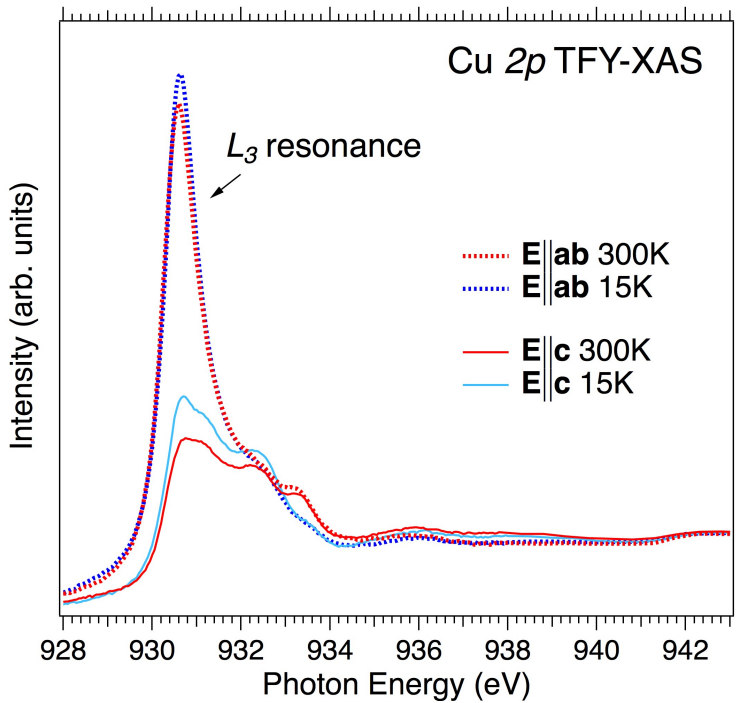

**Figure 1:** Polarization- and temperature-dependent X-ray absorption spectra of optimally doped YBCO at the Cu *2p* absorption edge.

### 2.2 Ligand-field multiplet calculations

The Cu $3d \rightarrow 2p$ RIXS spectra of YBCO were calculated in a coherent second-order optical process using the Kramers-Heisenberg formula [8] including interference effects in the intermediate states. The Slater integrals, describing the $3d$ - $3d$ and the $3d$ - $2p$ Coulomb and exchange interactions, and the spin-orbit splitting were obtained by the Hartree-Fock method [9]. The effect of the configuration-dependent hybridization was taken into account by scaling the Slater integrals to $F_k(3d3d)$ 80%, $F_k$ (2p3d) 80% and $G_k$ (2p3d) 80%. In a crystalline field of octahedral symmetry, the ground state is $^2E_g$ and in tetrahedral D$_{4h}$ symmetry, this $^2E_g$ state is further split into the $^2A_{1g}$ and $^2B_{1g}$ states while the $^1A_{1g}$ multiplet is the Zhang-Rice singlet. In $D_{4h}$ symmetry, the ground state of the $Cu^{2+}$ ion has $^2D_{5/2}$ character. Here, we calculate the many-electron ground state (multiplet) for $Cu^{2+}$ ions (Cu $3d^9$) in both $D_{4h}$ and $C_{4h}$ symmetries.

Due to the Jahn-Teller distortion, the $CuO_5$ pyramids are elongated, resulting in a longer Cu-O bond (~2.41 Å) along the *c*-axis and four shorter (~1.89 Å) Cu-O bonds in the basal plane. As the Cu electron configurations fluctuate between $3d^{10}$, $3d^9$ and $3d^8$ in the ground state, as in the case of binary transition metal oxides [10] [11], the wave function contains all three configurations in addition to several charge-transfer configurations that were considered.

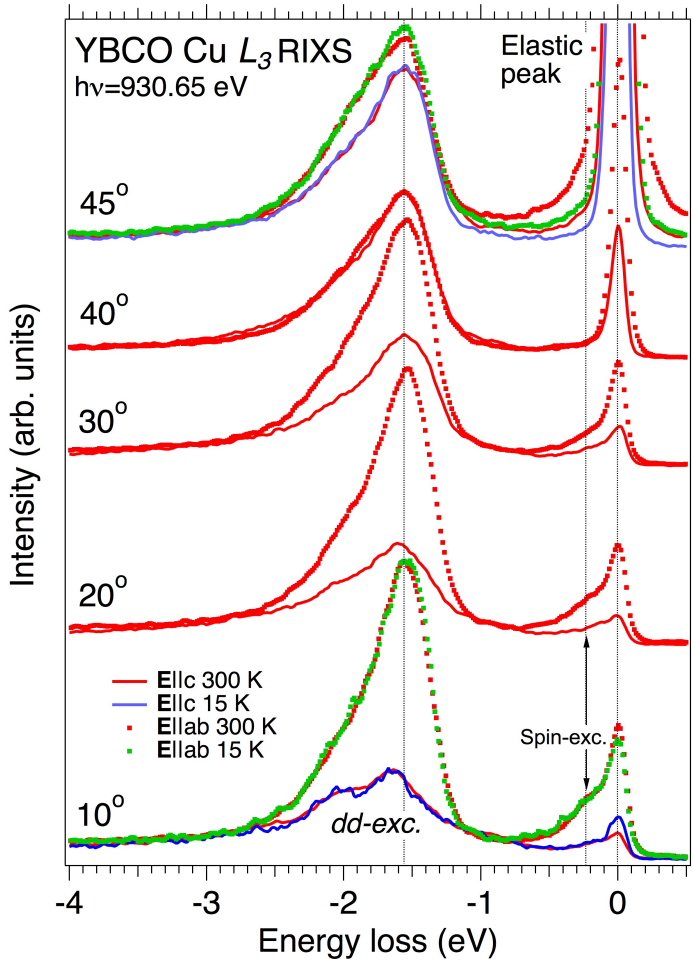

**Figure 2:** Angular-dependent Cu $L_3$-RIXS spectra at 10°, 20°, 30°, 40° and 45° grazing incidence angles with horizontal and vertical polarization for optimally doped YBCO.

The single-impurity Anderson model (SIAM) [12] with full multiplet effects was applied to describe the system. The crystal field and exchange interactions were taken into account by using a code by Butler [13] and the charge-transfer effect was implemented with a code [14] developed by Theo Thole. The charge-transfer energy $\Delta$, defined as the energy difference between the center of gravity between the main line and the different charge-transfer configurations, was chosen to reproduce the experiment ($\Delta_{gs}$=4.1 eV and $\Delta_{is}$=4.25 eV), and the crystal-field splitting in the $3d^9$ $D_{4h}$ local symmetry was $10Dq$=1.5, $D_s$=0.3, $D_t$=0.16. The three $E_{dxy}$=-1.50 eV, $E_{dxz}$ and $E_{dyz}$=-1.60 eV orbitals indicated in Fig. 3 are collectively referred to as $t_{2g}$ and the $E_{dx2-y2}$=0 and $E_{dz2-r2}$=-2.00 eV orbitals are referred to as $e_g$. The calculations were made for the same geometry as the experimental one, with the scattering angle between the incoming and outgoing photons fixed to 90°. The z-axis was perpendicular to the *ab* plane and the $d_{x2-y2}$ orbital was in the ab basal plane. To test the effect of the exchange interaction inducing spin-flip magnon excitations, the calculations were also made in the $C_4$ basis set with an inter-atomic exchange field of 0.21 eV.

## 2. Results and discussion

### 2.1 Cu *L*-edge XAS

Figure 1 shows incident angle- and temperature-dependent X-ray absorption spectra of optimally doped YBCO at the Cu *2p* absorption edge. In particular, we note the peculiar threshold shoulders that are absent in XAS spectra of pure Cu [15]. These shoulders are related to Zhang-Rice [3] plane-, and chain- excitations, respectively. The temperature-dependence of these shoulders have been thoroughly discussed in a previous paper [4]. In this paper, the energy at the main $L_3$ resonance is used as excitation energy for the *L*-edge RIXS spectra.

### 2.2 Cu *L*-edge RIXS

Figure 2 shows angular- and temperature-dependent Cu $L_3$-RIXS spectra of optimally doped YBCO measured at 10°, 20°, 30°, 40° and 45° (grazing) incidence angles with horizontal and vertical polarization for a scattering angle of 90°. At 45° incidence angle (specular geometry), the elastic peak is very intense, while at the lower angles, it is largely suppressed, in particular, when **E**||**c**. At 45° incidence angle, the intensity of the *dd*-excitations is already rather independent of the incident X-ray polarization while at the lower incidence angles, the intensity is very anisotropic with more unoccupied states in-plane than out-of-plane. Note that as the incidence angle is lowered, a shoulder below the elastic peak increases (-0.21 eV), in particular, for the in-plane excitation. This feature can tentatively be assigned to a local spin-flip magnon excitation while collective excitations such as acoustic phonon modes also occur in the mid-infrared region (MIR). Recent momentum dependent Cu *L*-RIXS results [16] [17] have revealed a considerable dispersion within this energy region that is mainly due to single magnon excitation, although other contributions are not negligible. In RIXS at the Cu *L*-edge of the $La_{2-x}Sr_xCuO_4$ family [18], of $Bi_2Sr_2CaCu_2O_{8+\delta}$ [19] as well as of Sr$_{14}$Cu$_{24}$O$_{41}$ [20] dispersive magnetic modes have been observed at similar energies. Here we will attempt to describe the Cu *L*-RIXS within a ligand field multiplet model [21] that project the local contributions.

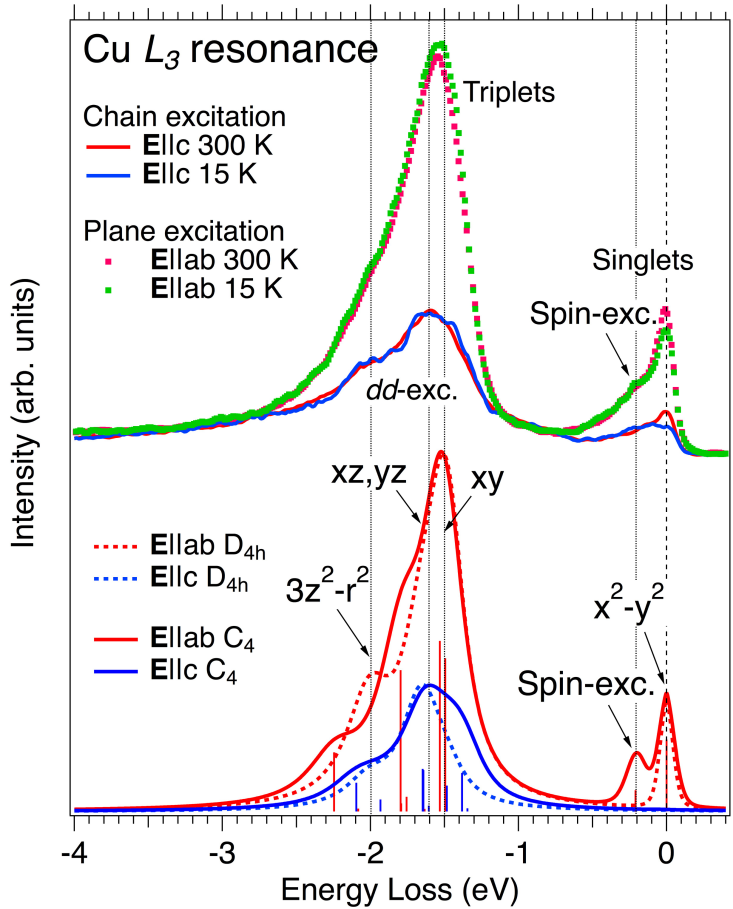

**Figure 3:** (top) Cu $L_3$ resonance RIXS spectra excited at 930.7 eV. Bottom: Calculated spectra with $3d^9$ ($Cu^{2+}$) final states in $D_{4h}$ and $C_4$ symmetry including Cu $3d_{x2-y2}$ super-exchange energy of 0.21 eV. The vertical blue and red sticks indicate the intensity and position of the transitions to the final states in the simulated RIXS spectra in $C_4$ symmetry. The three $E_{dxy}$=-1.50 eV, $E_{dxz}$ and $E_{dyz}$=-1.60 eV orbitals collectively referred to as $t_{2g}$ and the $E_{dx2-y2}$=0 and $E_{dz2-r2}$=-2.00 eV orbitals as $e_g$, are indicated by the vertical dotted lines.

Figure 3 shows a close up of the resonant *L$_3$* spectra measured at 20° incidence angle in comparison to calculated model spectra with *3d$^9$* (Cu$^{2+}$) final states in *D$_{4h}$* and *C$_4$* symmetries. The C$_4$ symmetry is applied to include a Cu *3d$_{x2-y2}$* super-exchange energy of 0.21 eV. Within the approximations of multiplet theory, there is no effective difference between the C$_{4h}$ and the C$_{4v}$ symmetries. This means that the crystal field parameters (10Dq, Ds and Dt) for D$_{4h}$ equally describe C$_{4v}$ and C$_{4h}$ symmetries. In undoped YBa$_2$Cu$_3$O$_6$, the Cu(1) and Cu(2) atoms in the chains and planes have *D$_{4h}$* and *C$_{4v}$* symmetries, respecively [22]. Thus, these symmetries are also appoximations of the doped YBa$_2$Cu$_3$O$_{7-\delta}$ system, where the Cu(1) and Cu(2) atoms in the chains and planes, respectively, have *D$_{2h}$* and *C$_{2v}$* symmetries. Excitation close to the *L$_3$* resonance (930.7 eV with 120 meV resolution) yields simulated spectra that are dominated by *dd*-excitations (of *3d$^9$*-configuration) around -1.5 eV. The Cu$^{2+}$ charge transfer multiplet final state orbital energies were then found to have energies of distorted divalent Cu orbitals *in* (E$_{dx2-y2}$=0, E$_{dxy}$=-1.50 eV) and *perpendicular* to (E$_{d3z2-r2}$=-2.00 eV) the CuO$_2$-planes. Based on our observations and experience from reasonable values of distortions in other cuprate systems [16] [23], the dx2-y2, dxy and dz2 orbital energies and the spin-flip excitation resolved in the RIXS experiment was transferred into crystal field parameters [14] (10Dq, Ds and Dt and the exchange field (0.21 eV). The effect of the exchange field in C$_4$ symmetry is clearly observed in the presented theoretical spectrum as a single magnon spin flip. The general spectral shape of the *dd*-orbital excitations as well as the spin-excitation peak is similar above and below the MST [14] [24]. Therefore, the residual temperature dependence in the experimental spectra is ascribed to effects on the collective lattice modes due to the MST. For instance, it is observed that the elastic peak loses intensity by cooling. Upon cooling, the intensity at 0 eV is lowered as the number of holes is increased and the superconducting gap is formed. As the carriers lower their kinetic energy at the MST, the spectral weight is transferred from the *E$_F$* and the elastic peak decreases. Recently, Magnuson *et al.* [4] were able to detect that cooling from room temperature to 15 K leads to a significant orbital reconstruction and charge transfer between the CuO$_2$ planes and the CuO$_3$-chains. In particular, the involvement of apical oxygen ions in the binding to the Cu ions in the planes and chains makes the temperature dependence of (non-electronic) excitations in the MIR intriguing.

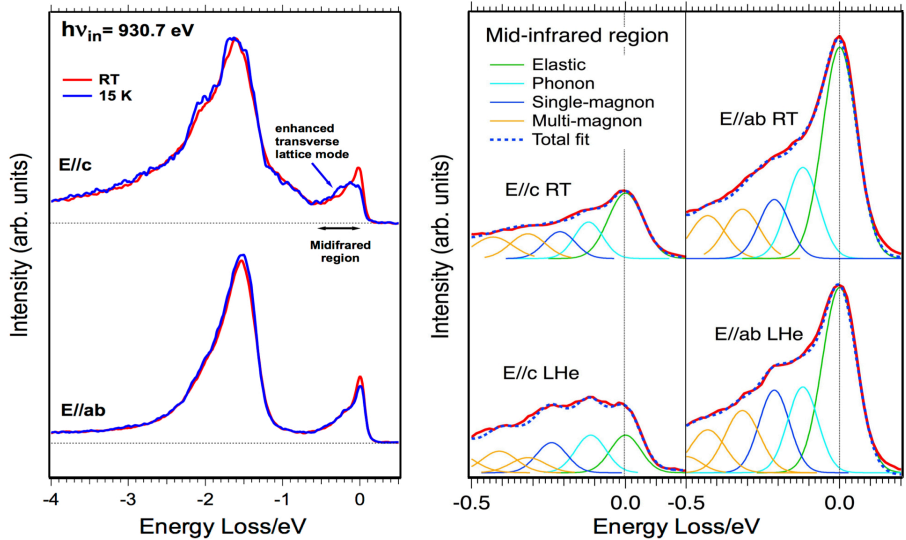

**Figure 4:** Mid-infrared energy region of RIXS excited at the Cu *L$_3$* resonance (930.7 eV). The elastic peak appears at 0 eV. Our curve fit helps identify a single phonon mode at 116 meV, a single-magnon mode at 230 meV (246 meV out-of-plane) and multi-magnon modes at 320 meV (mainly in-plane) and 390 meV.

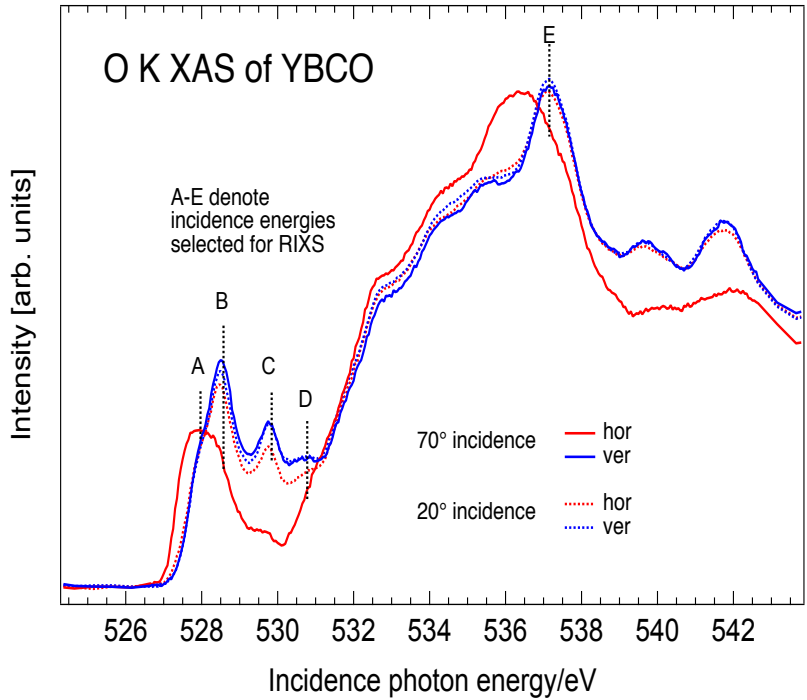

**Figure 5:** O $K$ XAS of optimally doped YBCO for two different incident polarizations.

Figure 4 (left panel) shows the temperature dependence of the resonantly excited Cu $L_3$ RIXS spectra. The arrow in the top of the left panel points out an enhancement (~250 meV) that the (**E**//**c**) lattice mode perpendicular to the CuO$_2$ basal plane acquires at low temperature. The right panel shows a close-up of the MIR for the same spectra. By using a phenomenological multi-Gaussian fit for **E**//**c** (left) and **E**//**ab** polarizations (right) we try to capture the general behaviour instead of the more rigorous approach for instance in Refs. [16] [23] [25]. This leads to a series of peaks that are narrower than use of the full formalism would yield. Nevertheless, they correspond to visible spectral features, and using the assignments of previous works as guide this allows us to make the following tentative assignments: the elastic peak at 0 eV, a possible single phonon mode at 116 meV (not reported in previous literature), a single-magnon mode at 230-246 meV (higher value out-of-plane) and multi-magnon modes at above 320 meV. This can be compared with similar values of Sr$_2$CuO$_2$Cl$_2$, where one-, two, and three-magnon contributions have been predicted around 300-400 meV [26] [27]. The higher energy and frequency of the single-magnon mode along the $c$-axis is due to a longer bond length between the Cu(1)O-chain orbitals compared to the in-plane Cu(2,3)O$_2$-plane orbitals in the basal $ab$-plane.

We find that sample cooling leads to a strong increase of the single magnon mode (dark blue traces in the left panels of Fig. 3) excited in the basal CuO$_2$ plane and a decrease of the out-of-plane excited elastic peak (green traces in the left panels of Fig. 3). Note that the phonon modes along the $c$-axis, in particular, show intriguing temperature dependences that will be addressed using future higher-resolving RIXS instrumentation. In addition, the elastic peak consists of optical phonons, yet unresolved by RIXS that could be important for the MST of YBCO. Recently, resonantly driven phonon modes by 20 THz infrared laser pulses in X-ray diffraction pump-probe experiments [28] have indicated a pico-second short transient superconducting state above room-temperature in YBCO. The Cu(1)-O(4) stretching mode ($A_g$ symmetry) couple to the transverse mode ($B_{1u}$ symmetry) [29] of the apical oxygen atoms between the planes and the chains that is involved in the self-doping process. However, the path to a continuous superconducting state at room temperature is hampered by the lack of fundamental understanding what drives the metal-superconductor transition (MST) in the first place.

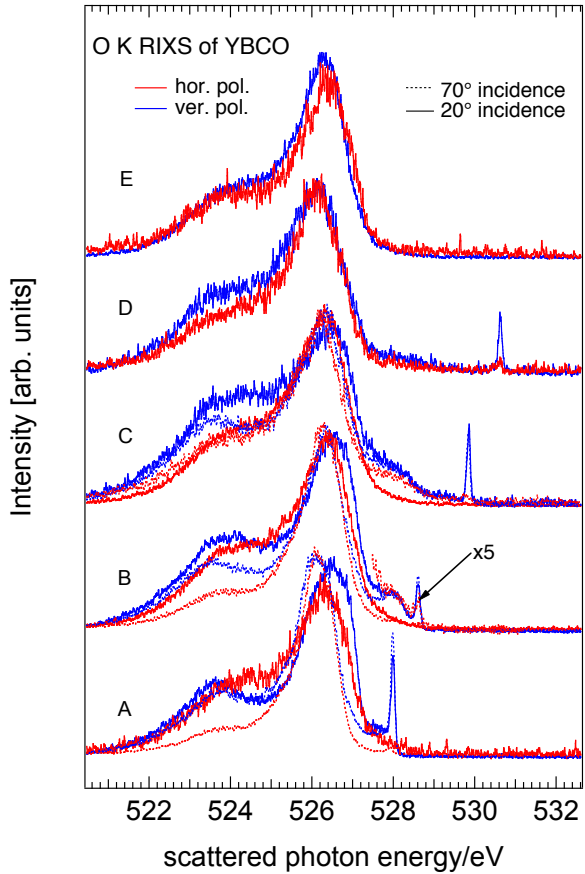

**Figure 6:** Polarization dependent RIXS excited with incident energies (denoted A-E) as depicted in Fig. 4.

### 2.3 O *1s*-edge XAS

The O *1s* edge XAS and O *K* RIXS spectra of YBCO give additional clues about the nature and temperature dependence of the electronic structure of high temperature superconductors. Previously, using O *1s* XAS, Merz et al. [30] estimated the hole distribution for $x=6.91$ in the $CuO_2$ plane, at the apical O and in the Cu-O chain as 0.40, 0.27 and 0.24, respectively. Figure 5 displays high resolution polarization dependent O *K*-XAS for optimally doped YBCO [31]. The crystal anisotropy (**E**||**c** vs **E**||**ab**) is clearly reflected by the polarization differences in the O *K* edge spectra as well as at the Cu *L*-edge. Horizontally and vertically polarized synchrotron radiation was used for two different incidence angles, 20° and 70°. For vertical polarization, the polarization vector of the incident X-rays **E**, is parallel to the $CuO_2$-planes, independent of the incidence angle. For the horizontal polarization, the polarization vector of the incident X-rays is dominantly in plane (out of plane) for 20° (70°). This means that the polarization vector only has a significant out of plane component in the configuration of horizontal polarization at an incidence angle of 70°. This is observed in Fig. 5 as the solid red trace that is significantly different from the other three traces for which the polarization vector is almost in plane. The pre-peak region, where the O *1s* electron is excited into the (doping-induced) Zhang-Rice [3] band (ZRB) and the Upper Hubbard band (UHB), respectively, is of main interest since these states involve hybridization with the correlated Cu *3d*-orbitals. The UHB (denoted C in Fig. 5) is highly polarized in plane, while there is additional intensity on the low energy shoulder of ZRB (A and B in Fig. 5) when the incident polarization vector is almost parallel with (the out of plane) crystal axis *c*. This feature (A), corresponds to the Cu(1)-O(4) apical oxygen and chain excitations [31].

### 2.4 O *K*-edge RIXS

Figure 6 displays polarization dependent RIXS spectra (using 70 meV combined resolution) with incident energies in the region of the ZRB and the UHB. Note that the incidence angle is a near-normal 20° while the exit angle is 70° due to the 90° scattering geometry. In this geometry, we observe a main band (525-527 eV) with a low-energy shoulder (522-525 eV) nearly stationary in photon energy that we can assign to O *2p*-states that are more or less covalently hybridized with Cu *3d*-orbitals, depending on the excitation energy. The Cu state are *dd*-excitations that show a strong polarization dependence at the lowest incident energy (A) [32] [33], corresponding to the inequivalent O(4) apical oxygen and O(1) chain sites. Previous O *K* RIXS studies with **E**||**ab** suggested that O(2,3) with *x*=6.94 have higher binding energy than those of O(1) and O(4) [34]. At 70° incidence angle, the main O *2p* peak is more localized at 525.2 eV emission energy but self-absorption effects [15] are more significant than at 20° incidence angle where this effect can be neglected. Furthermore, an interesting feature resonating at the ZRB peak B is observed in vertical polarization both for 20° and 70° incidence angle. This feature has a loss energy of less than 1 eV and only appears when the incident X-ray polarization vector is oriented in the *ab*-basal plane of the YBCO crystal. This observation is reminiscent of the bi-magnons observed in O *K* RIXS for e.g. of $La_xSr_{1-x}CuO_4$ observed by Bisogni *et al.* [35] apart from the relatively high energy loss excitation of more than 0.5 eV that we record for YBCO.

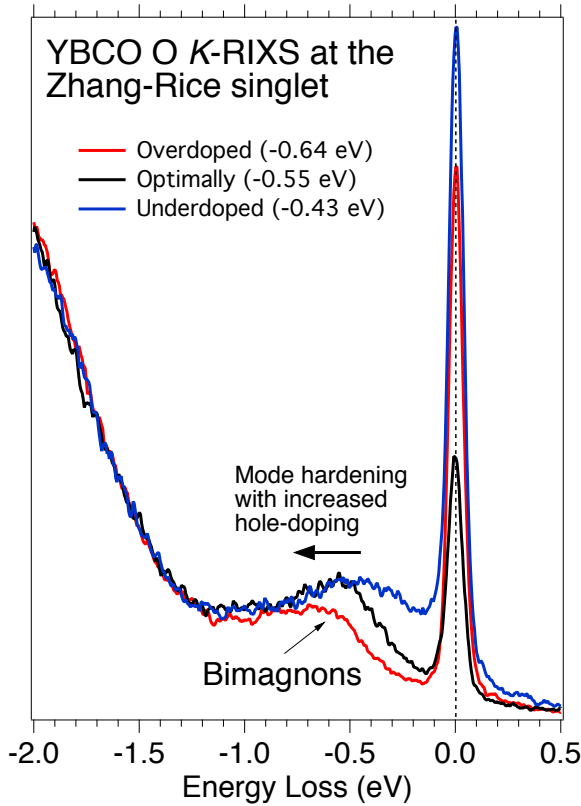

**Figure 7:** Doping dependent O *K*-RIXS that shows how the collective excitations harden for higher oxygen doping levels.

No ZRB RIXS feature appears for (nearly) out of plane (**E**||**c**) polarization, i.e. horizontal incident polarization of **E** together with a 70° incidence angle of the x-rays on the sample (red solid trace). For a 20° incidence angle (in the present 90° scattering geometry), the ratio between the oxygen main band and the ZRB RIXS feature is about 5 times as large as for vertical polarization (we therefore provide a magnification of this part of the spectrum in Fig. 6 with a corresponding scaling factor, see arrow). This observation emphasizes the completely different origins of these two spectral features. It is likely due to the two-dimensional character of the involved orbitals that prevents the suppression of intensity in this geometry for the oxygen main band feature.

We now turn the discussion of the possible origin of the ZRB RIXS feature. From Cu *L* RIXS (Fig. 2) we know that the single magnon has an energy around 0.2 eV, so that bi-magnons are allowed to appear at about double the energy, i.e. ~0.4 eV. We have also investigated the doping dependence of these excitations and plotted the result in Fig. 7. Note that there is a shift to higher loss energy (i.e. a *hardening*) with increasing oxygen doping, which is unexpected if it were an ordinary bi-magnon excitation [35] [36]. The hardening of this mode as doping increases is *opposite* to the behavior of the magnetic excitation spectra observed at the Cu *L*-resonance of the YBCO system [20]. Also at the Cu *K* and *L*-edges magnons are found to shift towards lower energy with increasing hole doping. For the LaSrCuO-system, one has only found rather little *q*-dependence [35], which was rationalized as being the effect of momentum averaging and thus reflecting the high-energy density of magnon states. A *q*-dependence has not yet been reported for bi-magnons so far but we have preliminary indications that this feature also has a momentum dependent behavior. The recent discovery of a strongly dispersive high energy mode by Ishii *et al.* [17] has been attributed to charge excitations. These are not limited in energy and low-energy orbital excitations might involve both Cu and O in the chains. Our finding of a relatively intense spectral feature in the mid IR region is indeed peculiar to YBCO, as LSCO and the quasi 1D cuprates do not show any intensity in that energy range. The doping-dependence is also puzzling with hardening of the excitation with increasing hole doping. However, the dispersion behaviors are markedly distinct and for YBCO the momentum dependent behavior is wave-like. Recently Lee *et al.* [37] resolved site-dependent low-energy excitations arising from progressive emissions of a 70 meV lattice vibrational mode in $Ca_{2+5x}Y_{2-5x}Cu_5O_{10}$. The dispersion of bimagnons are not yet fully understood and only some guiding theory exists [38]. We tentatively assign these excitations to collective magnetic excitations within the ZR and/or the Hubbard bands. However, further work is needed to assign this feature more definitely, as the assignment to complex collective magnetic excitations is not well understood theoretically. A forthcoming publication will discuss these excitations in more depth [39].

## 3. Summary and Conclusions

To summarize, we present high resolution XAS and RIXS results at the Cu *L* and O *K*-edge of $YBa_2Cu_3O_{6.9}$. Ligand-field multiplet calculations describe the observed *dd*-excitations very well and a spin flip excitation is found to be consistent with Cu *L* RIXS at medium momentum transfer. No temperature dependence could be observed for this mode, which suggests that it might not be directly involved in the MST. Finally, a novel doping and incident energy-dependent feature at around 0.5-0.6 eV is observed in O *K* RIXS of YBCO. Intriguingly, the mode hardens unexpectedly with oxygen doping. This excitation is interpreted as a multi-magnon excitation within the ZR- and/or Hubbard bands of YBCO.

## 4. Acknowledgements

We would like to thank the staff at the SLS for experimental support. This work was supported by the Swedish Research Council (VR) Linnaeus Grant LiLi-NFM, the FUNCASE project supported by the Swedish Strategic Research Foundation (SSF). MM acknowledges financial support from the Swedish Energy Research (no. 43606-1), *the Swedish Foundation for Strategic Research (SSF)* (no. RMA11-0029) *through the synergy grant FUNCASE* and the Carl Tryggers Foundation (CTS16:303, CTS14:310).